\documentclass[aps,amsmath,notitlepage,twocolumn,amssymb,prl,letterpaper,longbibliography]{revtex4-1}

\usepackage[pdftex]{graphicx} 
\usepackage{epstopdf}
\usepackage{verbatim}
\usepackage{color}
\usepackage{subfigure}
\usepackage{tabularx}
\usepackage{amsfonts}
\usepackage{wasysym}
\usepackage{bm}

\usepackage[colorlinks=true,citecolor=red,urlcolor=blue]{hyperref}

\definecolor{darkred}{rgb}{0.847059, 0.141176, 0.164706}
\definecolor{darkgreen}{rgb}{0,0.4,0}
\definecolor{darkblue}{rgb}{0.254902, 0.411765, 0.882353}

\newcommand{\be}{\begin{equation}}
\newcommand{\ee}{\end{equation}}

\newcolumntype{C}[1]{>{\centering\let\newline\\\arraybackslash\hspace{0pt}}m{#1}}

\begin{document}
\title{Symmetry Enriched U(1) Topological Orders for Dipole-Octupole Doublets on a Pyrochlore Lattice}
\author{Yao-Dong Li$^{1}$}
\author{Gang Chen$^{1,2}$}
\email{gangchen.physics@gmail.com}
\affiliation{$^{1}$State Key Laboratory of Surface Physics,
Center for Field Theory and Particle Physics,
Department of Physics, Fudan University, Shanghai 200433, People's Republic of China}
\affiliation{$^{2}$Collaborative Innovation Center of Advanced Microstructures,
Nanjing, 210093, People's Republic of China}

\date{\today}

\begin{abstract}
Symmetry plays a fundamental role in our understanding 
of both conventional symmetry breaking phases
and the more exotic quantum and topological phases
of matter. We explore the experimental signatures of 
symmetry enriched U(1) quantum spin 
liquids (QSLs) on the pyrochlore lattice. We point out that the Ce 
local moment of the newly discovered pyrochlore QSL candidate 
Ce$_2$Sn$_2$O$_7$, is a dipole-octupole doublet. The generic model 
for these unusual doublets supports two distinct symmetry enriched 
U(1) QSL ground states in the corresponding quantum spin ice regimes. 
These two U(1) QSLs are dubbed dipolar U(1) QSL and octupolar U(1) QSL. 
While the dipolar U(1) QSL has been discussed in many contexts, 
the octupolar U(1) QSL is rather unique.
Based on the symmetry properties of the dipole-octupole doublets,
we predict the peculiar physical properties of the octupolar U(1) QSL,
elucidating the unique spectroscopic properties in the external
magnetic fields. We further predict the Anderson-Higgs transition
from the octupolar U(1) QSL driven by the external magnetic fields.
We identify the experimental relevance with the candidate material
Ce$_2$Sn$_2$O$_7$ and other dipole-octupole doublet systems.
\end{abstract}

\maketitle

\textit{Introduction.}---The interplay between symmetry and topology is the frontier 
subject in modern condensed matter physics~\cite{ZCGu2009,Pollmann2012,Senthil2015}.
At the single particle level, the non-trivial realization of time reversal symmetry
in electron band structure has led to the discovery of topological insulators~\cite{hasan2010colloquium,qi2011topological}.
For the intrinsic topological order such as Z$_2$ toric code
and chiral Abelian topological order, a given symmetry of the system
could enrich the topological order into distinct phases that cannot be
smoothly connected without crossing a phase
transition~\cite{Wen200710, Ran2013,Hermele2013,Lu2016}.
Despite the active theoretical efforts, the experimentally relevant 
symmetry enriched topological order is extremely rare. 
In this work, we explore one physical realization
of {\sl symmetry enriched U(1) topological order} for the  
dipole-octupole (DO) doublets on the pyrochlore lattice
and predict the experimental consequences 
of distinct symmetry enrichment.
The DO doublet is a special Kramers' doublet in the
D$_{\text{3d}}$ crystal field environment~\cite{Chen2014,Yaodong2016,li2016hidden}.
Both states of the DO doublet transform as the
one-dimensional irreducible representations ($\Gamma_5^+$ or $\Gamma_6^+$)
of the D$_{\text{3d}}$ point group~\cite{Chen2014}. 
It was realized that the DO doublets on the pyrochlore lattice could support
two distinct U(1) quantum spin liquid (QSL) ground states~\cite{Chen2014}.
These distinct U(1) QSLs are the symmetry enriched
U(1) topological orders~\footnote{We here adopt the definition of the
U(1) topological order in Ref.~\onlinecite{Hermele2004}. 
The concept is more general than the fully gapped intrinsic topological 
orders that are described by topological quantum field theory.} 
and are enriched by the lattice symmetries of the pyrochlore systems.

\begin{figure}[tp]
\centering
\includegraphics[width=8cm]{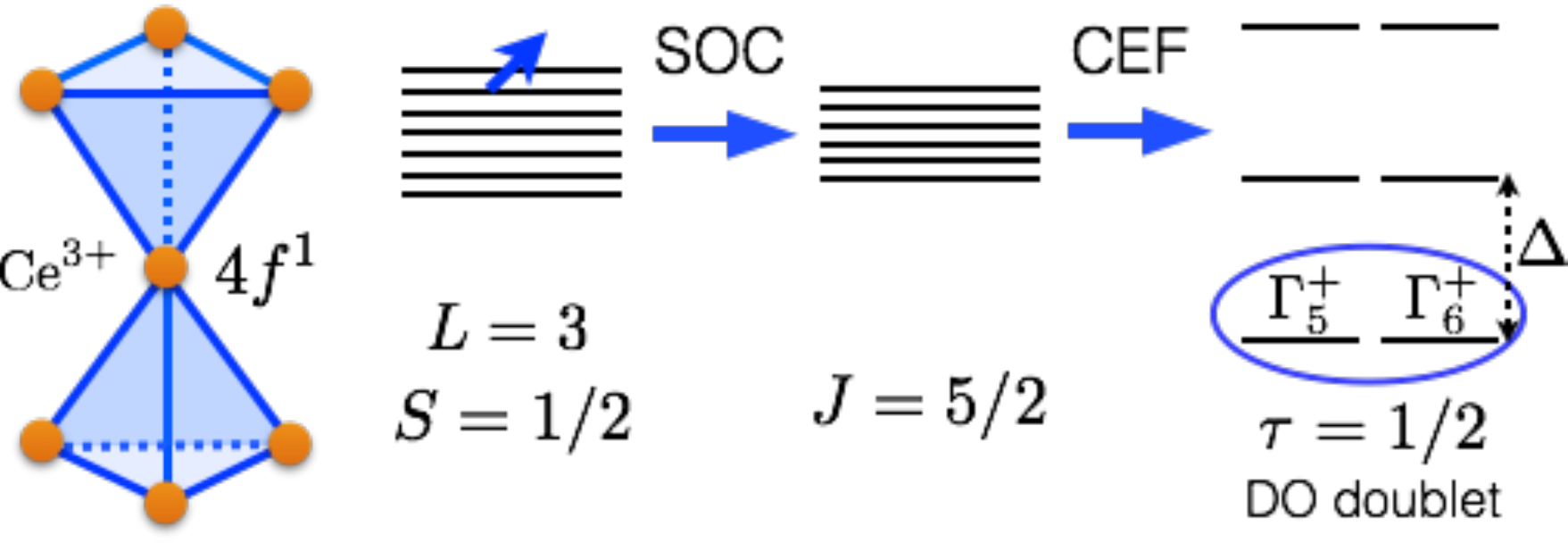}
\caption{The electron configuration and the D$_{\text{3d}}$
crystal electric field (CEF) splitting of the
Ce$^{3+}$ ion in Ce$_2$Sn$_2$O$_7$.
The CEF ground state wavefunctions are combinations
of $J^z = \pm 3/2$ states~\cite{Sibille2015}, thus the
CEF ground state is a DO doublet.
$\Delta$ is the CEF gap and was
fitted to be $\Delta = 50 \pm 5$meV~\cite{Sibille2015}.}
\label{fig1}
\end{figure}

Recently Ce$_2$Sn$_2$O$_7$ was proposed as the first Ce-based QSL 
candidate in the pyrochlore family~\cite{Sibille2015}, in which no 
magnetic order was observed down to 0.02K. 
Although it was not noticed previously, the Ce$^{3+}$ local moment 
in Ce$_2$Sn$_2$O$_7$ is actually a DO doublet. 
The strong atomic spin-orbit coupling (SOC) of the 4$f^1$ electron 
in the Ce$^{3+}$ ion entangles the electron spin ($S=1/2$) 
with the orbital angular momentum ($L=3$) 
into a $J=5/2$ total moment. The six-fold degeneracy of the $J=5/2$ 
total moment is further splitted into three Kramers' doublets by 
the D$_{\text{3d}}$ crystal field (see Fig.~\ref{fig1}). Since the ground state doublet 
wavefunctions are combinations of $J^z = \pm 3/2$ states~\cite{Sibille2015}, 
this doublet is precisely the DO doublet that we defined~\cite{Chen2014}. 
Because the crystal field gap is much larger than the interaction energy 
scale of the local moments and the temperature scale in the experiments, 
the low temperature magnetic property of Ce$_2$Sn$_2$O$_7$ is fully governed 
by the ground state doublets.

\begin{figure*}[th]
\centering
\includegraphics[width=.32\textwidth]{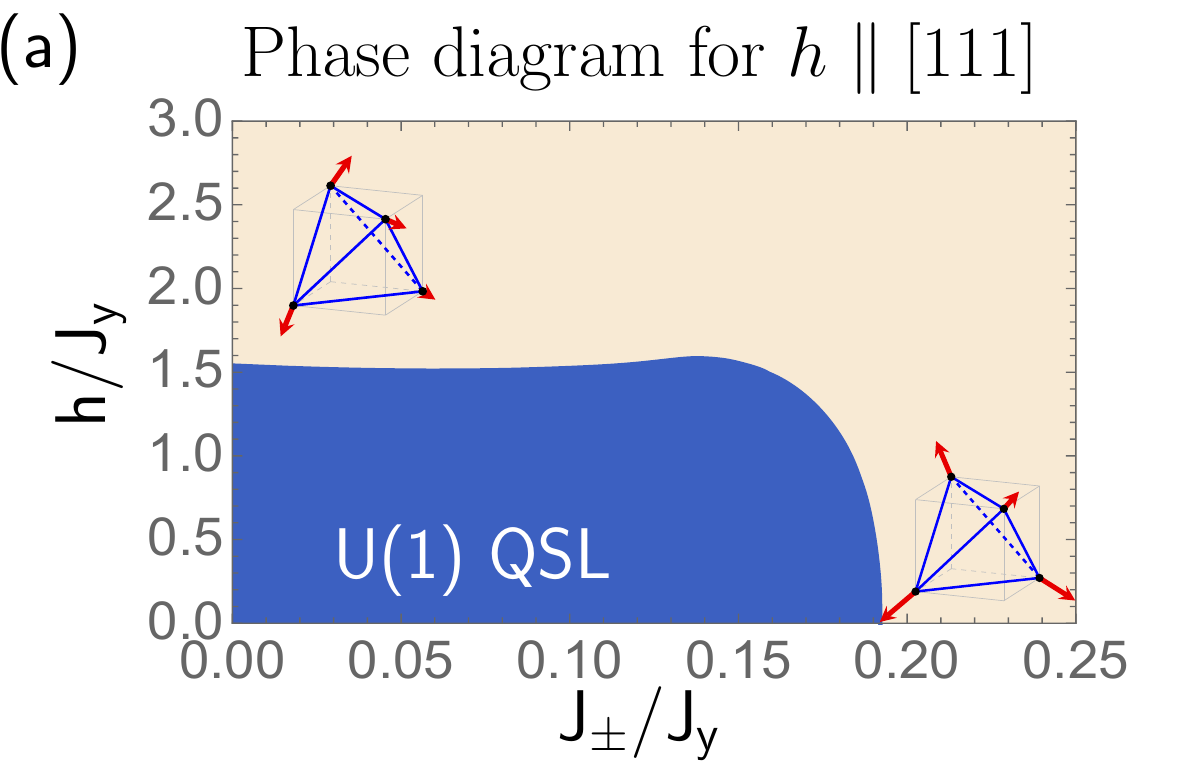}
\includegraphics[width=.32\textwidth]{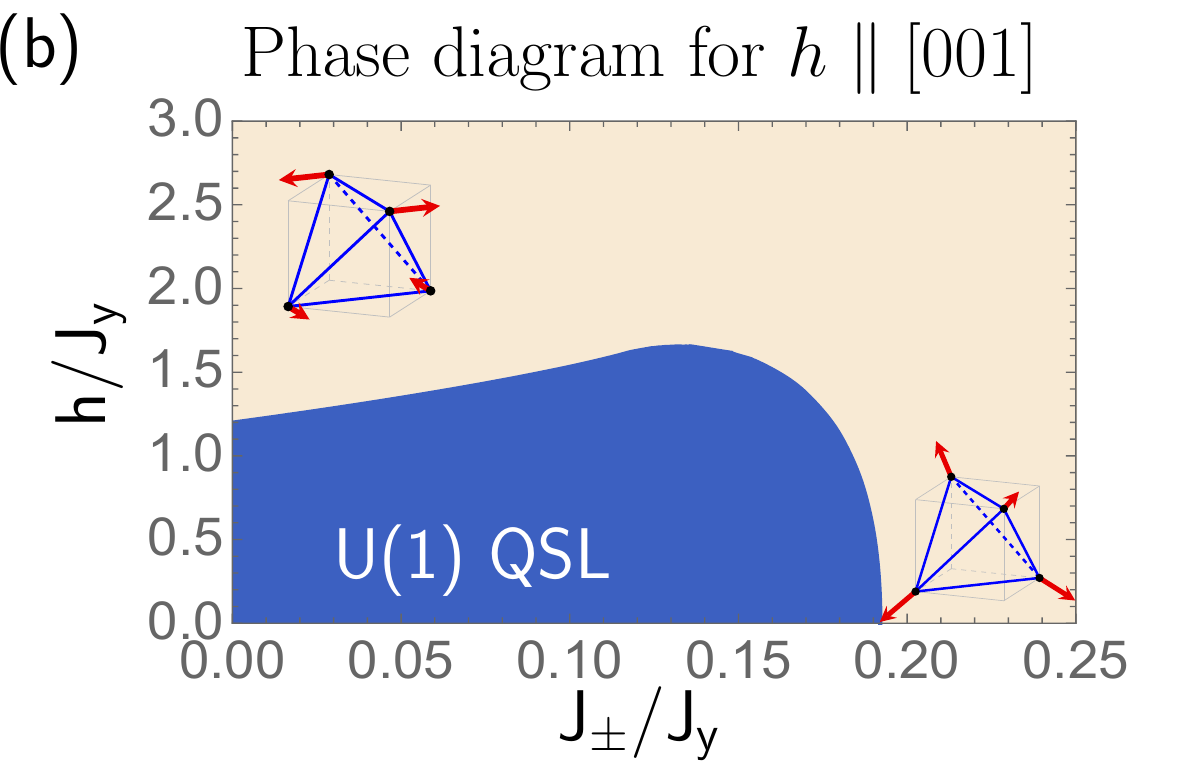}
\includegraphics[width=.32\textwidth]{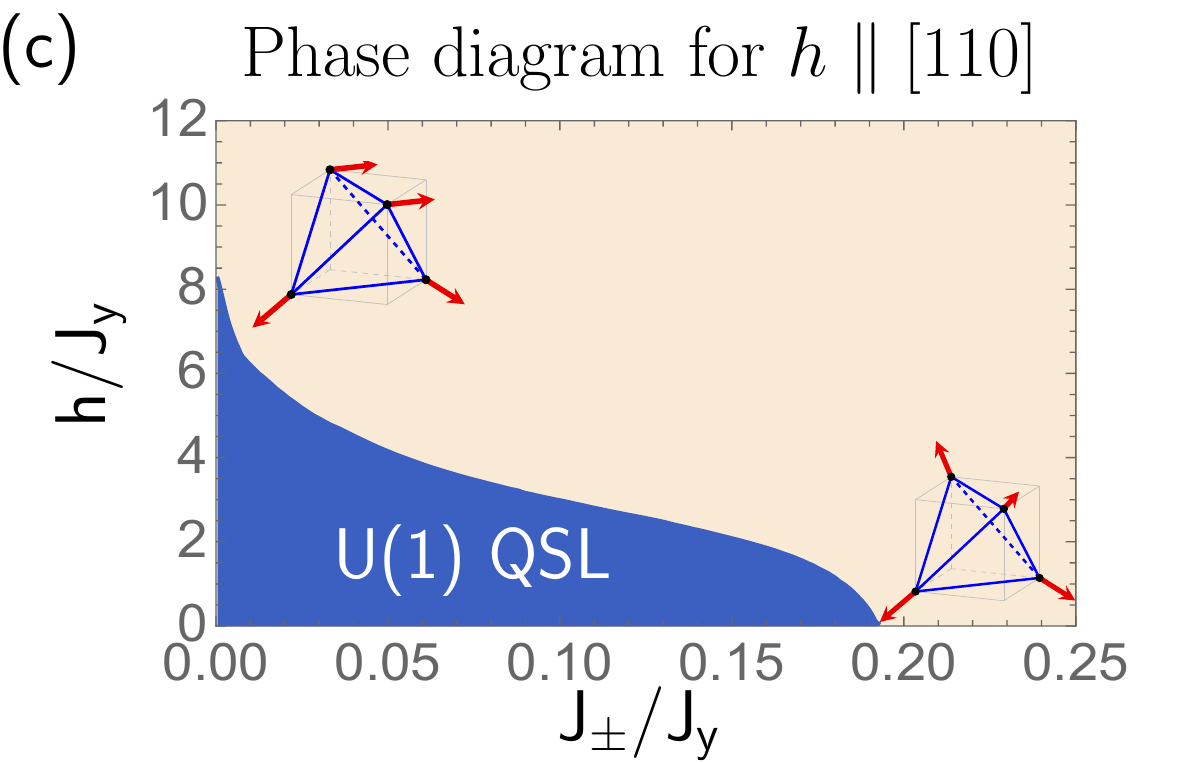}
\caption{
Phase diagrams for magnetic fields along (a) $[111]$,
(b) $[001]$, and (c) $[110]$ directions. Outside the
QSL phases are the induced magnetic ordered phase via the
spinon condensation. For $h=0$, the spinons are condensed
at ${\bf k}_c = (0,0,0)$, and we choose the local moments
to order in the local $\hat{z}$ direction. In (a), large
magnetic field near the vertical axis drives the spinon
condensation at ${\bf k}_c=\pi(1,1,1)$, and the resulting
order is depicted in the figure. This order smoothly connects
to the order on the horizontal axis. The cases in (b) and
(c) are similar, except that in (b) the field on the vertical
axis drives the condensation at ${\bf k}_c = 2\pi(0,0,1)$,
while in (c) ${\bf k}_c = \pi(1,1,0)$ near the vertical axis.
We set the diamond lattice constant to unity.
}
\label{fig2}
\end{figure*}

Motivated by the experiments on Ce$_2$Sn$_2$O$_7$
and more generally by the experimental consequences of
the distinct symmetry enriched U(1) QSLs for the DO doublets,
in this Letter, we explore the peculiar properties of
the DO doublets in external magnetic fields. In the octupolar
U(1) QSL of the octupolar quantum spin ice regime for
the DO doublets, we find that the external magnetic field
directly couples to the spinons and modifies the spinon dispersions.
This effect allows us to control the spinon excitations with
the magnetic fields. The lower excitation edge of the spinon continuum  
in the dynamic spin structure factors can thus be modified by the
magnetic fields, which gives a sharp prediction for
the inelastic neutron scattering experiments. When the magnetic
field exceeds the critical value and closes the spinon gap, 
the spinons are condensed, driving the system through an 
Anderson-Higgs' transition and inducing the long-range 
magnetic orders.

\textit{Generic model for DO doublets on the pyrochlore lattice.}---Because
of the peculiar symmetry properties of the DO doublets, the most generic
model that describes the nearest-neighbor interaction between them is given
as $H_{\text{DO}} = \sum_{\langle ij \rangle} [ J_x^{} \tau^x_i \tau^x_j
+ J_y^{} \tau^y_i \tau^y_j + J_z^{} \tau^z_i \tau^z_j
+ J_{xz}^{} (\tau^x_i \tau^z_j + \tau^z_i \tau^x_j) ] $~\cite{Chen2014}.
Here the interaction is uniform on every bond despite the fact that
the DO doublet involves a significant contribution from the orbital
part due to the strong SOC~\cite{Chen2008,Jackeli2009,Chen2010,
William14,Khaliullin2005,Chen2011}, and the DO doublet is modeled
by an effective pseudospin-1/2 moment ${\boldsymbol{\tau}}$.
Both $\tau^x$ and $\tau^z$ transform as the dipole moments under the space
group symmetry, while the $\tau^y$ component behaves as an octupole 
moment~\cite{Chen2014}. It is this important difference that leads 
to some of the unique properties of its U(1) QSL ground states.

Due to the spatial uniformity of the generic model, we
can transform the model
$H_{\text{DO}}$ into the XYZ model with
\begin{eqnarray}
H_{\text{XYZ}} = \sum_{ \langle ij \rangle }
  \tilde{J}_x^{}  \tilde{\tau}^x_i \tilde{\tau}^x_j
+ \tilde{J}_y^{}  \tilde{\tau}^y_i \tilde{\tau}^y_j
+ \tilde{J}_z^{}  \tilde{\tau}^z_i \tilde{\tau}^z_j ,
\end{eqnarray}
where $\tilde{\tau}^x$ and $\tilde{\tau}^z$ ($\tilde{J}_x$ and
$\tilde{J}_z$) are related to $\tau^x$ and $\tau^z$ (${J}_x$
and ${J}_z$) by a rotation around the $y$ direction in the pseudospin
space, and $\tilde{\tau}^y \equiv {\tau}^y, \tilde{J}_y \equiv {J}_y$.
When one of the couplings, $\tilde{J}_{\mu}$, is dominant
and antiferromagnetic, the corresponding pseudospin component,
$\tilde{\tau}^{\mu}$, is regarded as the Ising component of the model,
and the ground state is a U(1) QSL in the corresponding
quantum spin ice regime. The dipolar U(1) QSL is realized
when the Ising component is the dipole moment $\tilde{\tau}^x$
or $\tilde{\tau}^z$, while the octupolar U(1) QSL is realized
when the Ising component is the octupole moment $\tilde{\tau}^y$.
In the compact U(1) quantum electrodynamics description of the low energy
properties of the U(1) QSL~\cite{Hermele2004,PhysRevLett.91.167004}, the Ising component
is identified as the emergent electric field~\cite{Hermele2004}.
Therefore, the emergent
electric field transforms very differently under
the lattice symmetry in dipolar and octupolar U(1) QSLs, making these
two U(1) QSLs symmetry enriched U(1) topological order on the pyrochlore
lattice~\cite{Chen2014}.

\emph{Octupolar U(1) QSL and field-driven Anderson-Higgs' transitions.}---Since
the dipolar U(1) QSL has been discussed many times in literature~\cite{Savary2012,Sungbin2012,Chen2014,Zhihao2014,Moessner2015,
Chen2015,Shannon12,Chen2016,Shannon2012,Gingras2014},
we here focus on the octupolar U(1) QSL of the octupolar quantum spin
ice regime where $\tilde{J}_y$ is dominant and antiferromagnetic.
The octupolar U(1) QSL is a new phase that is {\sl unique} to the DO doublet
and cannot be found in any other doublets on the pyrochlore lattice.

We consider the coupling of the DO doublet to the external magnetic
field. Remarkably, because $\tilde{\tau}^y$ is an octupole moment,
it does not couple to the magnetic field even though it is time
reversally odd. Only the dipolar component, $\tau^z$, couples
linearly to the external magnetic field. The resulting model is
\begin{eqnarray}
H & = & \sum_{ \langle ij \rangle } \sum_{\mu = x,y,z}
  \tilde{J}_{\mu}^{}  \tilde{\tau}^{\mu}_i \tilde{\tau}^{\mu}_j
- \sum_i h \, (\hat{n} \cdot \hat{z}_i ) \, \tau^z_i,
\label{eq2}
\end{eqnarray}
where $\hat{n}$ is the direction of the magnetic field
and $\hat{z}_i$ is the $z$ direction of the local coordinate
basis at the lattice site $i$~\cite{Supple}.
This generic model describes {\sl all} magnetic properties
of the DO doublets on the pyrochlore lattice.

As the generic model contains four parameters, it necessarily
brings some unnecessary complication into the problem.
To capture the essential physics, we here consider
a simplified version of the generic model in Eq.~(\ref{eq2}).
The simplified model is
\begin{eqnarray}
H_{\text{sim}} & = & \sum_{\langle ij \rangle} J_y^{} \tau^y_i \tau^y_j
- J_{\pm}^{} (\tau^+_i \tau^-_j + h.c.)
\nonumber \\
&-& \sum_i h \, (\hat{n} \cdot \hat{z}_i ) \, \tau^z_i,
\label{Hsim}
\end{eqnarray}
where we define $\tau^{\pm}_i = \tau_i^z \pm i \tau_i^x$
and $\hat{n}$ is the direction of the external magnetic field.
In the Ising limit with $J_{\pm}=0$ and $h = 0$, the antiferromagnetic
$J_y$ favors the $\tau^y$ components to be in the ice manifold
and requires a ``two-plus two-minus'' ice constraint for the $\tau^y$
configuration on each tetrahedron. This octupolar ice manifold
is extensively degenerate. With a small and finite $J_{\pm}$ or $h$,
the system can then tunnel quantum mechanically within the octupolar
ice manifold and form an octupolar U(1) QSL.
In this perturbative limit, the degenerate perturbation theory yields
an effective ring exchange model with~\cite{Supple}
\begin{eqnarray}
H_{\text{ring}} =  J_{\text{ring}} \sum_{\hexagon}
\big[ \tau^+_i \tau^-_j \tau^+_k \tau^-_l \tau^+_m \tau^-_n + h.c. \big],
\end{eqnarray}
where ``$i,j,k,l,m,n$'' are six sites on the perimeter of the elementary
hexagon of the pyrochlore lattice, and the ring exchange
$J_{\text{ring}} < 0$ for $J_{\pm} > 0$ and for either sign of $h$.
$H_{\text{ring}}$ does not involve defect tetrahedra that violate
the ice constraint and thus only describes the quantum fluctuation
and dynamics {\sl within} the ice manifold. It is well-known that
the low energy properties of $H_{\text{ring}}$ is described by the
compact U(1) quantum electrodynamics~\cite{Hermele2004} of the U(1) QSL with
gapless gauge photon, and the spin-flip operator $\tau^{\pm}_i$
is identified as the gauge string within the ice manifold.
We expect the simplified model $H_{\text{sim}}$ captures
the generic properties of the octupolar U(1) QSL.

\begin{figure}[t]
\centering
\includegraphics[width=.235\textwidth]{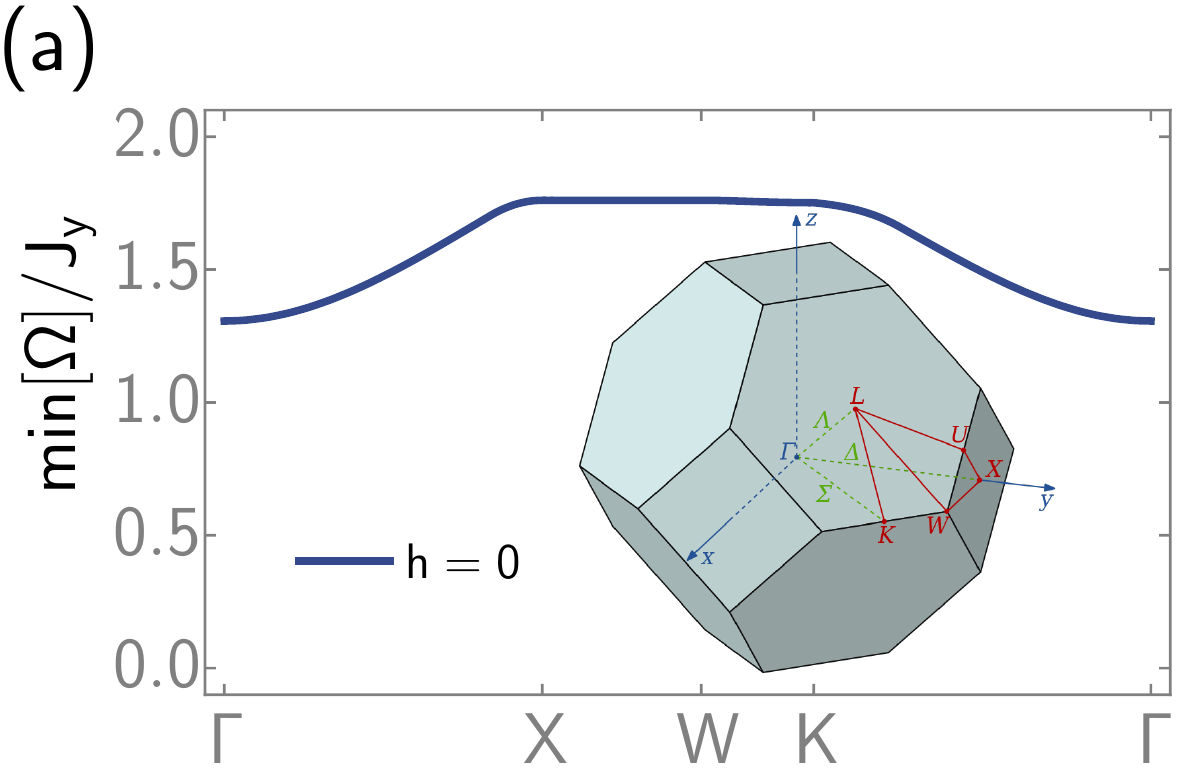}
\includegraphics[width=.235\textwidth]{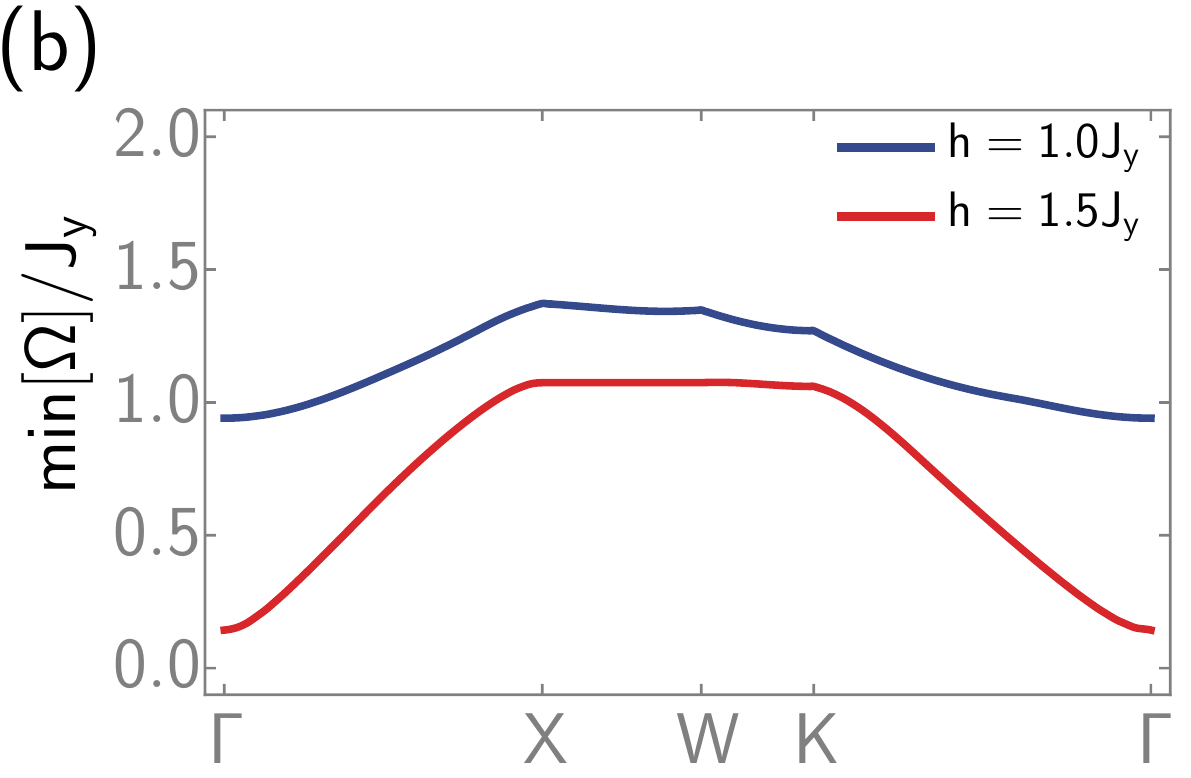}
\vspace{0.2cm}\\
\includegraphics[width=.235\textwidth]{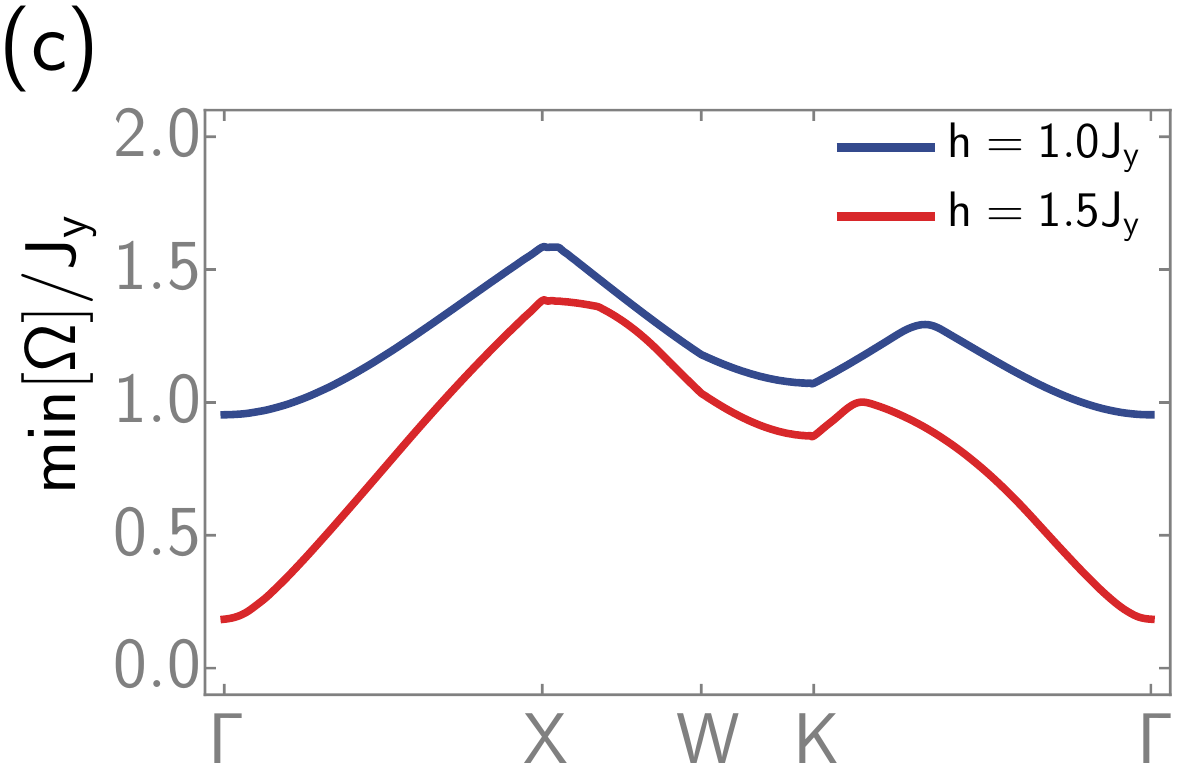}
\includegraphics[width=.235\textwidth]{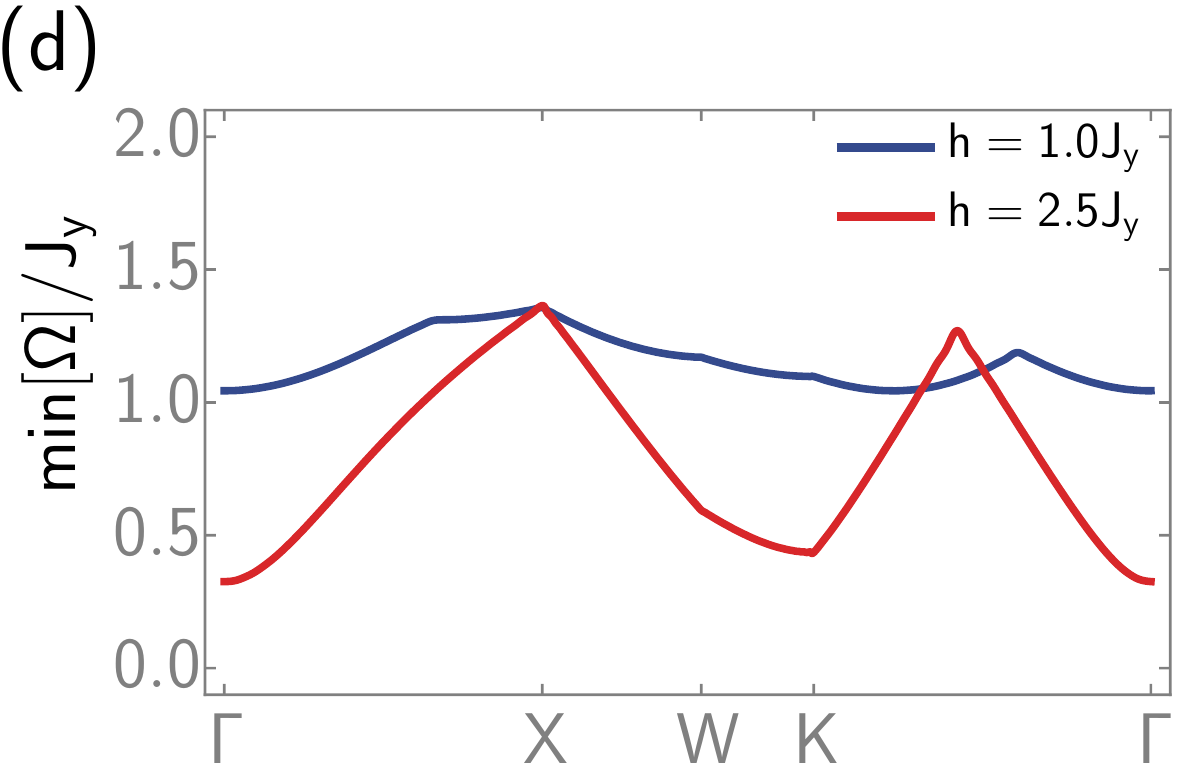}
\caption{Lower excitation edges of the spinon continuum in the
dynamic spin structure factor under (a) zero magnetic field, and field
along (b) [111], (c) [001], and (d) [110] directions..
In the figure, we set $J_{\pm} = 0.1 J_y$. The inset of (a) is
the Brillouin zone~\cite{wiki}.}
\label{fig3}
\end{figure}

To obtain the phase diagram of $H_{\text{sim}}$, we start
from the octupolar U(1) QSL phase and study its instability.
For this purpose, we include the spinon excitations
(that are out of the ice manifold) into the formulation.
The perturbative analysis and $H_{\text{ring}}$, that focus on the ice manifold,
does not capture the spinons. We here implement a parton-gauge
construction for the octupolar U(1) QSL and formulate $H_{\text{sim}}$
into a lattice gauge theory with the spinons.
Like many other parton construction, we replace the physical
Hilbert space with a larger one and supplement it with a constraint.
We follow Refs.~\onlinecite{Savary2012,Sungbin2012}
and express the pseudospin operators as
\begin{eqnarray}
\tau^+_i = \Phi^\dagger_{\bf r} \Phi^{}_{{\bf r}'} s^+_{{\bf r}{\bf r}'},
\quad\quad
\tau^y_i = s^y_{{\bf r}{\bf r}'},
\end{eqnarray}
where ${\bf r}{\bf r}'$ is the link that connects two neighboring
tetrahedral centers at ${\bf r}$ and ${\bf r}'$, and the pyrochlore
site $i$ is shared by the two tetrahedra. The centers of the tetrahedra
form a diamond lattice, and ${\bf r}$ (${\bf r}'$) belongs to the I (II)
diamond sublattice. Here ${\bf s}_{{\bf r}{\bf r}'}$ is a spin-1/2 variable
that corresponds to the emergent gauge field, and
$\Phi^{\dagger}_{\bf r}$ ($\Phi^{}_{\bf r}$) creates (annihilates)
one spinon at the diamond site ${\bf r}$.
The spinons carry the emergent electric charge,
and $\Phi^{\dagger}_{\bf r}$ and $\Phi^{}_{\bf r}$
are raising and lowering operators of the emergent electric charge.
Since we enlarged the physical Hilbert space, the constraint
${Q}_{\bf r} = \eta_{\bf r} \sum_{\mu} \tau^y_{{\bf r},
{\bf r} + \eta_{\bf r} {\bf e}_{\mu}} $
is imposed, where $\eta_{\bf r} = 1 \, (-1)$ for the I (II) sublattice
and the ${\bf e}_{\mu}$'s are the first neighbor vectors of the diamond lattice.
Here ${Q}_{\bf r}$ measures the electric charge at ${\bf r}$ and satisfies
\begin{eqnarray}
[\Phi_{\bf r}^{}, Q_{\bf r}^{}] = \Phi_{\bf r}^{}, \quad \quad
[\Phi^\dagger_{\bf r}, Q^{}_{\bf r}] = - \Phi^\dagger_{\bf r}.
\end{eqnarray}

The U(1) QSL of quantum spin ice is an example of the string-net
condensed phases~\cite{Levin2005}. In the U(1) QSL,
$\tau^{\pm}_i$ creates the shortest open (gauge) string whose ends
are spinon particles. In the spin ice context,
$\tau^{\pm}_i$ creates two defect tetrahedra that violate the ``two-plus
two-minus'' ice constraint. The parton-gauge construction captures this
essential property, and the model becomes
\begin{eqnarray}
H_{\text{sim}} &=& \sum_{\bf r} \frac{J_y Q_{\bf r}^2 }{2}
- \sum_{{\bf r}} \sum_{\mu \neq \nu} J_{\pm}
\Phi_{{\bf r}+ \eta_{\bf r}{\bf e}_{\mu}}^{\dagger}
\Phi_{{\bf r}+\eta_{\bf r} {\bf e}_{\nu}}^{}
s^{-\eta_{\bf r}}_{{\bf r},{\bf r}+\eta_{\bf r} {\bf e}_{\mu}}
\nonumber \\
&\times &
s^{+\eta_{\bf r}}_{{\bf r},{\bf r}+\eta_{\bf r}{\bf e}_{\nu}}
- \sum_{ \langle {\bf r}{\bf r}' \rangle}\frac{h}{2} (\hat{n} \cdot \hat{z}_i )
(\Phi^{\dagger}_{\bf r} \Phi^{}_{{\bf r}'} s^+_{{\bf r}{\bf r}'}+ h.c.).
\label{eqLGT}
\end{eqnarray}
With the constraint, Eq.~(\ref{eqLGT}) is an exact reformulation of
the simplified model in Eq.~(\ref{Hsim}). It describes the bosonic
spinons hopping on the diamond lattice.
The spinons are minimally coupled with the emergent U(1) gauge field.
Remarkably, the external magnetic field directly couples to the
spinons and does {\sl not} couple to the emergent electric field.
This is sharply distinct from the dipolar U(1) QSL where the magnetic
field would also directly couple with the emergent electric field.

Inside the U(1) QSL, the spinons are fully gapped. 
The external magnetic field allows the spinon to tunnel 
between the neighbor tetrahedra that are located along the field direction.
As we increase the magnetic field $h$, the spinon gap gradually
decreases. It is expected that, at a critical field strength,
the spinon gap is closed and the spinons are condensed with
$\langle \Phi_{\bf r} \rangle \neq 0 $.
Via the Anderson-Higgs' mechanism, the U(1) gauge field becomes
massive and gapped. Note this differs the Coulomb ferromagnet
where the gauge field remains gapless and deconfined~\cite{Savary2012}. 
The resulting proximate state develops
a long-range magnetic order. Therefore, this is an
Anderson-Higgs' transition driven by the external magnetic fields.
This is a generic property of the octupolar U(1) QSL and is not
a specific property of the simplified model.
To our knowledge, this is the first example that an
external probe drives an Anderson-Higgs' transition
in a physical system.

\begin{table*}[htp]
\begin{tabular}{ccc}
\hline\hline
Different U(1) QSLs    & Heat capacity & Inelastic neutron scattering measurement
\\
Octupolar U(1) QSL for DO doublets & $C_v \sim T^3$ & Gapped spinon continuum
\\
Dipolar U(1) QSL for DO doublets & $C_v \sim T^3$ & Both gapless gauge photon and gapped spinon continuum
\\
Dipolar U(1) QSL for non-Kramers' doublets~\cite{Sungbin2012} & $C_v \sim T^3$ & Gapless gauge photon
\\
Dipolar U(1) QSL for usual Kramers' doublets~\cite{Savary2012} & $C_v \sim T^3$ & Both gapless gauge photon and gapped spinon continuum
\\
\hline\hline
\end{tabular}
\caption{List of the physical properties of different U(1) QSLs
on the pyrochlore lattice. ``Usual Kramers doublet'' refers to
the Kramers doublet that is not a DO doublet.
They transform as a two-dimensional irreducible representation
under the D$_{\text{3d}}$ point group.
Although the dipolar U(1) QSL for DO doublets behaves the same
as the one for usual Kramers' doublets, their physical origins
are rather different~\cite{Supple}. }
\label{table1}
\end{table*}

To solve the reformulated model in Eq.~(\ref{eqLGT}), we adopt the
gauge mean-field approximation~\cite{Savary2012,Sungbin2012,Chen2014,Zhihao2014}.
In this approximation, we decouple
the model into the spinon sector and the gauge sector. Since
$H_{\text{ring}}$ favors a zero background gauge flux on each
elementary hexagon of the diamond lattice, we solve for the
mean-field ground state within this sector~\cite{Supple}.
The magnetic dipolar order is obtained by evaluating
\begin{eqnarray}
\langle \tau^z_i \rangle &=&
\frac{1}{2} \big[ \langle \tau^+_i \rangle  +  \langle \tau^-_i \rangle  \big]
\\
&=&\frac{1}{2} \big[
                            \langle \Phi^\dagger_{\bf r}  \Phi^{}_{{\bf r}'} \rangle
                            \langle s^+_{{\bf r}{\bf r}'} \rangle
                             + h.c.\big],
\end{eqnarray}
where $\langle \cdots \rangle$ is taken with respect to the ground state.
Because of the Zeeman coupling, $\langle \tau^z_i \rangle$ is non-zero
even in the U(1) QSL phase where the spinons are not
condensed. In the proximate ordered state, the spinon condensate
gives an additional contribution that is the induced magnetic order.
For all three directions of the external magnetic field, even though
the spinons are condensed at finite momenta, the proximate magnetic
order preserves the translation symmetry.

The full phase diagrams and the field-induced proximate
magnetic orders are depicted in Fig.~\ref{fig2}.
The magnetic field is found to be least effective in destructing
the U(1) QSL for the field along the [110] direction.
This is because the local $\hat{z}$ direction of two sublattices
are orthogonal to the [110] direction and the pseudospins on them
do not couple to the external field.
The phase transition is found to be continuous within the gauge
mean-field theory and may turn weakly first order after the
fluctuations are included. Nevertheless, as the spinon gap is
very small near the phase transition, this means that the heat
capacity and the magnetic entropy will be more pronounced at
low temperatures in these regions.

\textit{Lower excitation edges of the dynamic spin structure factors.}---
A smoking gun confirmation of U(1) QSL is to directly measure the gapless
U(1) gauge photon and/or the spinon continuum by inelastic neutron
scattering (INS) measurement. For the DO doublet, the neutron spin
couples to the local moment in the same way as the external magnetic field.
Therefore, for the octupolar U(1) QSL, the INS directly probes the
spinon excitation, and one would {\sl only} observe the spinon continuum
instead of the gapless U(1) gauge photon. The latter was proposed for
the dipolar U(1) QSL. This is the sharp difference between the octupolar
U(1) QSL and the dipolar U(1) QSL.

In the U(1) QSL, the spinon excitation has two branches due to the
two sublattice structure of the diamond lattice. {\sl Specifically} for the
simplified model $H_{\text{sim}}$, the two spinon branches are
degenerate in the absence of the external magnetic field
because the spinons do not hop from one sublattice to another.
As shown in Eq.~(\ref{eqLGT}), however, the magnetic field allows the spinons
to tunnel between the sublattices and breaks
the degeneracy of the two spinon bands. The splitted spinon bands are
labeled by $\omega_{1}({\bf k})$ and $\omega_{2}({\bf k})$~\cite{Supple}.

The INS measures the dynamic spin structure factor
$\langle \tau^z \tau^z \rangle_{{\bf q},\Omega}$, where ${\bf q}$
and $\Omega$ are the neutron momentum and energy transfer, respectively.
As $\tau^z$ is a spinon bilinear, one neutron spin flip creates
one spinon-antispinon pair that shares the neutron energy and
momentum transfer. From the conservation of the momentum and
the energy, we have
\begin{eqnarray}
{\bf q} &=& {\bf k}_1 + {\bf k}_2,
\\
{\Omega}({\bf q}) &=& \omega_i ({\bf k}_1) + \omega_j ({\bf k}_2),
\end{eqnarray}
where $i,j = 1,2$ are the band indices, and
${\bf k}_1$ and ${\bf k}_2$ are the momenta of the two spinons.

The lower excitation edge of the dynamic spin structure factor
encodes the minimum of the spinon excitation $\Omega({\bf q})$
for each ${\bf q}$. In Fig.~\ref{fig3}, we plot the dispersion
of the lower spinon excitation edge along the high symmetric
momentum direction in the octupolar U(1) QSL for different
external field orientations. The field modifies the spinon
dispersion and then tunes the spinon excitation edge.
As far as we are aware of, this is a rare example that one
can control the spinon excitations in a QSL.

\textit{Discussion.}---Many DO doublet pyrochlores are
actually magnetically ordered~\cite{NdHfO,NdZrO1,NdSnO,NdZrO2,NdZrO3,NdZrO4,CdErO,CdYbS},
which makes the QSL candidate Ce$_2$Sn$_2$O$_7$ rather unique.
Ce$_2$Sn$_2$O$_7$ has the Curie-Weiss
temperature $\Theta_{\text{CW}} \approx -0.25$K.
It was argued in Ref.~\onlinecite{Sibille2015}
that an antiferromagnetic $\Theta_{\text{CW}}$ cannot
support a QSL in the spin ice regime.
This conclusion is certainly true for the usual Kramers' doublet,
but is not the case for the DO doublets.
For the DO doublets, what $\Theta_{\text{CW}}$ measures is $J_z$,
not $\tilde{J}_z$ nor $\tilde{J}_x$~\cite{Supple}.
What determines the phase diagram of $H_{\text{XYZ}}$ 
are $\tilde{J}_{\mu}$'s, not the sign or value of
the single parameter $J_z$. One cannot rule out the possibility 
of the dipolar U(1) QSL in Ce$_2$Sn$_2$O$_7$. Moreover, 
the occurrence of octupolar U(1) QSL as a ground state of $H_{\text{XYZ}}$
is actually insensitive to the sign of $J_z$. 
If the ground state of Ce$_2$Sn$_2$O$_7$
does not belong to any other QSLs, the question then 
nails down to whether it is a dipolar U(1) QSL or an 
octupolar U(1) QSL.

In Tab.~\ref{table1} we list the thermodynamic and spectroscopic
properties of various U(1) QSLs.
Clearly, thermodynamic measurements cannot differentiate
them because the low-energy properties are all described
by the compact U(1) quantum electrodynamics. The INS
measurement, however, is a powerful technique to
identify the dipolar U(1) QSL and the octupolar U(1) QSL for the DO doublets.
As we wrote in Tab.~\ref{table1}, the INS can observe
both spinon continuum and gapless gauge photon for the dipolar U(1) QSL
while only gapped spinon continuum can be detected for the octupolar U(1) QSL.
We further propose the field driven Anderson-Higgs' transition and
the field-controlled dynamic spin structure factor
as the unique signatures of the octupolar U(1) QSL.
All these prediction can be useful to identify 
the nature of the QSL ground state in Ce$_2$Sn$_2$O$_7$.

To summarize, we predict a field driven Anderson-Higgs' transition
of the octupolar U(1) QSL for the dipole-octupole doublets on
the pyrochlore lattice. Inside the U(1) QSL, the lower excitation
edges of the spinon continuum are manipulated by the external
magnetic fields. This result provides a detectable experimental 
consequence in the INS measurements. We expect our work will 
surely stimulate the experimental studies of Ce$_2$Sn$_2$O$_7$ 
and other pyrochlore systems with dipole-octupole doublets.

\textit{Acknowledgements.}---This work is supported by the Start-up
funds of Fudan University (Shanghai, People's Republic of China)
and the Thousand-Youth-Talent program of People's Republic of China.

\bibliography{refsDOdoublet}

\newpage

\appendix

\begin{widetext}
\vspace{0.5cm}
{\Large Supplementary Information for ``Symmetry Enriched U(1) Topological Orders for Dipole-Octupole Doublets on a Pyrochlore Lattice''}
\vspace{0.5cm}
\end{widetext}

\section{I. Local coordinates and the generic model}

The local coordinate system at each sublattice of the pyrochlore lattice is defined in 
Tab.~\ref{local_coord}. 
\begin{table}[h]
\begin{tabular}{c c c c c}
  \hline
  \hline
  $\mu$ & 0 & 1 & 2 & 3 \\
  $\hat{{x}}_{\mu}$ & $\frac{1}{\sqrt{2}}[\bar{1}10]$ & $\frac{1}{\sqrt{2}}[\bar{1}\bar{1}0]$ & $\frac{1}{\sqrt{2}}[110]$ & $\frac{1}{\sqrt{2}}[1\bar{1}0]$ \\
  $\hat{{y}}_{\mu}$ & $\frac{1}{\sqrt{6}}[\bar{1}\bar{1}2]$ & $\frac{1}{\sqrt{6}}[\bar{1}1\bar{2}]$ & $\frac{1}{\sqrt{6}}[1\bar{1}\bar{2}]$ & $\frac{1}{\sqrt{6}}[112]$ \\
  $\hat{{z}}_{\mu}$ & $\frac{1}{\sqrt{3}}[111]$ & $\frac{1}{\sqrt{3}}[1\bar{1}\bar{1}]$ & $\frac{1}{\sqrt{3}}[\bar{1}1\bar{1}]$ & $\frac{1}{\sqrt{3}}[\bar{1}\bar{1}1]$ \\
  \hline
  \hline
\end{tabular}
\caption[Table caption text]{The local coordinate systems 
for the four sublattices of the pyrochlore lattice.}
\label{local_coord}
\end{table}

The dipole moment $\tau^z$ is defined in the local $\hat{z}$ direction, while 
the other two components $\tau^x$ and $\tau^y$ are defined in the pseudospin space. 
The magnetization of the system is thus given by 
\begin{equation}
{\bf m} = g\mu_{\text{B}}\sum_{i} \tau^z_i \, \hat{z}_i,
\end{equation}
where $g$ is the Land\'{e} factor and $\mu_{\text{B}}$ is Bohr magneton. 

To transform $H_{\text{DO}}$ 
($H_{\text{DO}} = \sum_{\langle ij \rangle}
[ J_x^{} \tau^x_i \tau^x_j + J_y^{} \tau^y_i \tau^y_j + J_z^{} \tau^z_i \tau^z_j
+ J_{xz}^{} (\tau^x_i \tau^z_j + \tau^z_i \tau^x_j) ]$)
to $H_{\text{XYZ}}$, we perform a rotation in the pseudospin space 
around the local-$y$ axis,
\begin{eqnarray}
{\tau}^x & = &  \cos\theta\,\tilde\tau^x + \sin\theta\, \tilde\tau^z, \\
{\tau}^y & = & \tilde\tau^y,  \\
{\tau}^z & = & {-\sin\theta }\, \tilde\tau^x + \cos\theta\, \tilde\tau^z ,
\end{eqnarray}
where $\tan 2\theta = {2J_{xz}}/({J_z - J_x})$.
Correspondingly,
\begin{eqnarray}
    \tilde{J}_y &=& J_y, \\
    \tilde{J}_x &=& \frac{1}{2}\left(J_x + J_z - \sqrt{4J_{xz}^2+(J_x-J_z)^2} \right), \\
    \tilde{J}_z &=& \frac{1}{2}\left(J_x + J_z + \sqrt{4J_{xz}^2+(J_x-J_z)^2} \right).
\end{eqnarray}

\section{II. Curie-Weiss temperatures}

Since the magnetization ${\bf m}$ is only related to the dipole moment $\tau^z$, 
the Curie-Weiss temperature only detects the interaction between $\tau^z$. 
From the original model $H_{\text{DO}}$, we carry out the high temperature
series expansion and find that
\begin{eqnarray}
\Theta_{\text{CW}} = + \frac{J_z}{2}. 
\end{eqnarray}
$\Theta_{\text{CW}}$ does not depend the orientation of the external probing field.

\section{III. Perturbation theory} 

Here we discuss the perturbation theory of the simplified model $H_{\text{sim}}$ 
with
\begin{eqnarray}
H_{\text{sim}} & = & \sum_{\langle ij \rangle} J_y^{} \tau^y_i \tau^y_j 
- J_{\pm}^{} (\tau^+_i \tau^-_j + h.c.)
\nonumber \\                    
&-& \sum_i h \, (\hat{n} \cdot \hat{z}_i ) \, \tau^z_i.
\end{eqnarray}
In the perturbative limit where $h \ll J_y$ and $J_{\pm} \ll J_y$, 
we carry out the degenerate perturbation theory to obtain the ring 
exchange interaction within the ice manifold. 

\begin{figure}[tp]
\centering
\includegraphics[width=6.0cm]{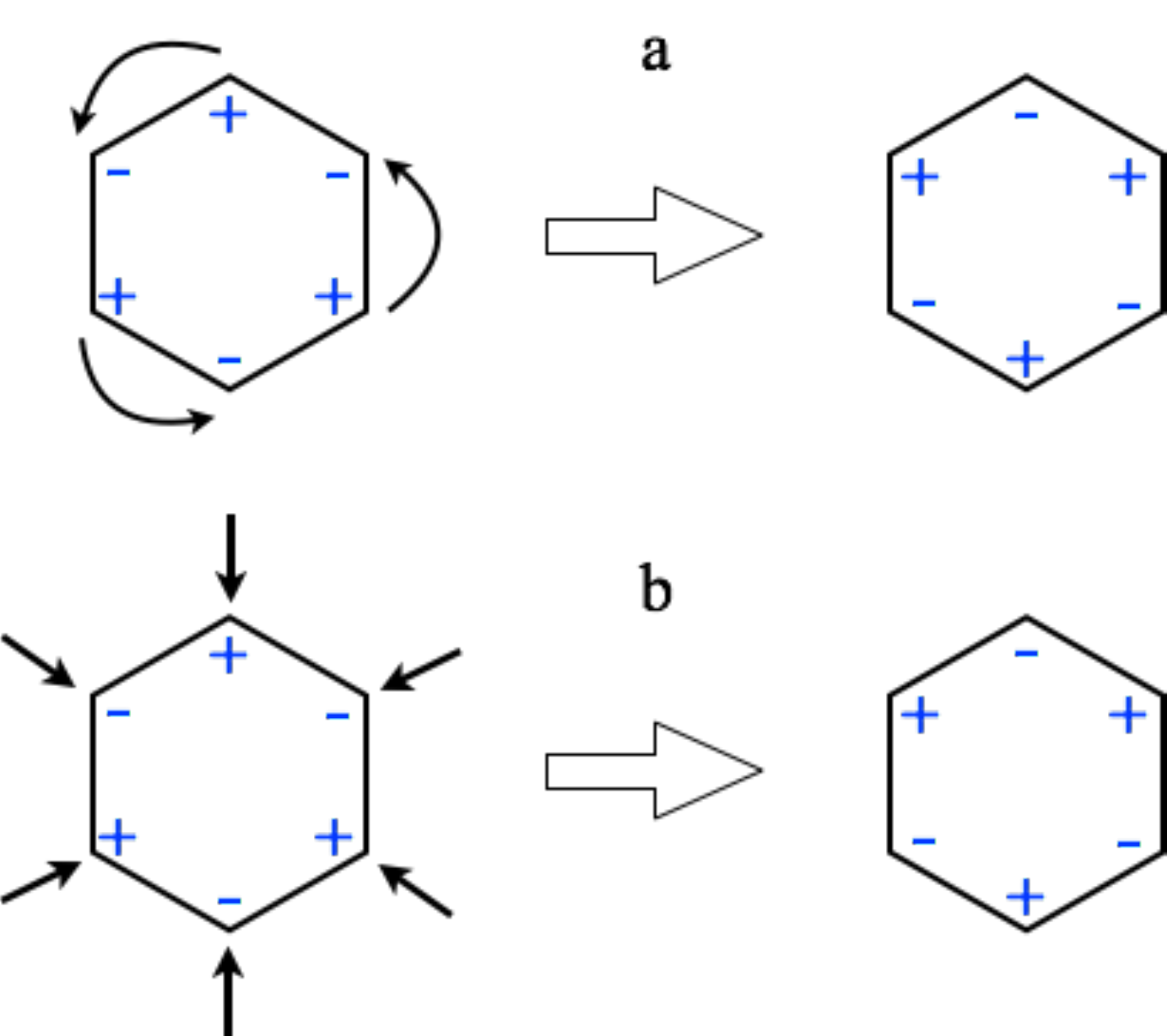}
\caption{The degenerate perturbation process
on the elementary hexagon of the pyrochlore lattice. 
Here ``$+$'' and ``$-$'' represent the orientation 
of the $\tau^y$ direction. (a) The curved arrows 
represents applying $J_{\pm}( \tau^+_i \tau^-_j + h.c.)$ 
on the bond. (b) The straight arrows represents applying 
$h \tau^z_i$ on the site.}
\label{sfig1}
\end{figure}

Without the external magnetic field, it is well-known that a third order 
degenerate perturbation is needed to generate the ring exchange 
(see Fig.~\ref{sfig1}a). Without the $J_{\pm}$, we need a sixth order degenerate 
perturbation of the external magnetic field to create quantum tunneling within 
the octupolar ice manifold (see Fig.~\ref{sfig1}b). When both the external field 
and the $J_{\pm}$ terms are present, the degenerate perturbation would always 
involve both $J_{\pm}$ and $h$ to generate the ring exchange. 
Therefore, in the ring exchange model, 
\begin{eqnarray}
H_{\text{ring}} = J_{\text{ring}} \sum_{\hexagon} 
\big[ \tau^+_i \tau^-_j \tau^+_k \tau^-_l \tau^+_m \tau^-_n + h.c. 
\big],
\end{eqnarray}
the coupling $J_{\text{ring}}$ has the following expression,
\begin{equation}
J_{\text{ring}} = 
\sum_{n_1,n_2}  
C_{n_1,n_2} h^{n_1} (-J_{\pm})^{n_2}  ,
\label{Jring}
\end{equation} 
where $C_{n_1,n_2}$ is a numerical coefficent 
in the perturbation series and $n_1$ is always even. 
The latter is because applying the Zeeman term 
one time only flips $\tau^y$ once. To get back to the ice manifold, 
we must always apply the Zeeman term even number of times. 
If the total perturbation order $n_1+n_2$ is even (odd),   
$C_{n_1,n_2}$ must be negative (positive). For a positive 
$C_{n_1,n_2}$, if $J_{\pm} > 0$, then every term in 
$J_{\text{ring}}$ gives a negative contribution and 
$J_{\text{ring}} < 0$. 
Precisely for the same reason, the simplified model
$H_{\text{sim}}$ does not have a sign problem for 
quantum Monte Carlo for $J_{\pm} > 0$ and for either sign
of $h$. 

Since the pseudospin operators $\tau^{\pm}_i$ in $H_{\text{ring}}$
are restricted to the spin ice manifold, we then can reexpress 
$\tau^{\pm}$ as 
\begin{equation}
\tau^{\pm}_i \simeq e^{\pm i A_{{\bf r}{\bf r}'}}
\end{equation}
where ${\bf r}$ and ${\bf r}'$ are the centers of the two neighboring tetrahedra
of the pyrochlore lattice site $i$, ${\bf r} \in $ I sublattice and ${\bf r}' \in $ 
II sublattice, and the $2\pi$ periodic phase variable $A_{{\bf r}{\bf r}'}$ satisfies 
$A_{{\bf r}{\bf r}'} = - A_{{\bf r}'{\bf r}}$. After this transformation, the ring
exchange becomes
\begin{eqnarray}
H_{\text{ring}} \simeq 2 J_{\text{ring}} \sum_{\hexagon_d} \cos (\text{curl} \, A  ) ,
\end{eqnarray}
where $\text{curl} \, A $ is the lattice curl on the elementary hexagon ($\hexagon_d$)
of the diamond lattice formed by the centers of the pyrochlore tetrahedra. 
Since $J_{\text{ring}} <0$, the ground state favors a zero flux 
with $\text{curl} \, A = 0$ on each hexagon 
of the diamond lattice. 

\section{IV. Gauge mean-field theory}

To implement the gauge mean-field theory~\cite{Savary2012,Sungbin2012,Chen2014,Zhihao2014}, 
we decouple the spinon-gauge coupling in the 
reformulated lattice gauge Hamiltonian of the main text
into the spinon sector and the gauge 
sector. The decoupling procedure is given as follows,
\begin{eqnarray}
&&
  \Phi^{\dagger}_{{\bf r} +\eta_{\bf r} e_{\mu} }   
  \Phi^{\phantom\dagger}_{{\bf r} +\eta_{\bf r} e_{\nu} }   
s^{-\eta_{\bf r}}_{{\bf r},{\bf r}+\eta_{\bf r} e_{\mu}}
s^{+\eta_{\bf r}}_{{\bf r},{\bf r}+\eta_{\bf r} e_{\nu}} \rightarrow
\langle s^{-\eta_{\bf r}}_{{\bf r},{\bf r}+\eta_{\bf r} e_{\mu}} \rangle
\langle s^{+\eta_{\bf r}}_{{\bf r},{\bf r}+\eta_{\bf r} e_{\nu}} \rangle 
\nonumber \\
&&
\quad\quad\quad\quad \times \big[
 \Phi^{\dagger}_{{\bf r} +\eta_{\bf r} e_{\mu} }   \Phi^{\phantom\dagger}_{{\bf r} +\eta_{\bf r} e_{\nu} }    - \langle 
\Phi^{\dagger}_{{\bf r} +\eta_{\bf r} e_{\mu} }   \Phi^{\phantom\dagger}_{{\bf r} +\eta_{\bf r} e_{\nu} }   \rangle
\big]
\nonumber 
\\
&&
\quad\quad\quad\quad + \big[
\langle s^{-\eta_{\bf r}}_{{\bf r},{\bf r}+\eta_{\bf r} e_{\mu}} \rangle
s^{+\eta_{\bf r}}_{{\bf r},{\bf r}+\eta_{\bf r} e_{\nu}} 
+ s^{-\eta_{\bf r}}_{{\bf r},{\bf r}+\eta_{\bf r} e_{\mu}}
\langle s^{+\eta_{\bf r}}_{{\bf r},{\bf r}+\eta_{\bf r} e_{\nu}}  \rangle
\nonumber \\
&&
\quad\quad\quad\quad -
\langle s^{-\eta_{\bf r}}_{{\bf r},{\bf r}+\eta_{\bf r} e_{\mu}} \rangle
\langle s^{+\eta_{\bf r}}_{{\bf r},{\bf r}+\eta_{\bf r} e_{\nu}}  \rangle
\big]
\langle
  \Phi^{\dagger}_{{\bf r} +\eta_{\bf r} e_{\mu} }   
  \Phi^{\phantom\dagger}_{{\bf r} +\eta_{\bf r} e_{\nu} }   \rangle ,
\\
&& 
\Phi^{\dagger}_{\bf r} \Phi^{}_{{\bf r}'} s^+_{{\bf r}{\bf r}'}
\rightarrow 
\big[ \Phi^{\dagger}_{\bf r} \Phi^{}_{{\bf r}'}  
- \langle \Phi^{\dagger}_{\bf r} \Phi^{}_{{\bf r}'}  \rangle \big]
\langle s^+_{{\bf r}{\bf r}'} \rangle 
+  \langle \Phi^{\dagger}_{\bf r} \Phi^{}_{{\bf r}'}  \rangle
 s^+_{{\bf r}{\bf r}'} .
\end{eqnarray}

With the above decoupling, the gauge sector is trivially solved 
for the zero gauge flux sector, and we have $s \equiv \langle 
s^{\pm}_{{\bf r}{\bf r}'}\rangle = 1/2$ on every link of the 
diamond lattice. For the spinon sector, the spinon mean-field 
Hamiltonian is now reduced to
\begin{eqnarray}
H_{\text{spinon}} &=&  \frac{J_y}{2} \sum_{\bf r} Q_{\bf r}^2 
                     - J_{\pm} s^2 
                      \sum_{{\bf r}} \sum_{\mu \neq \nu} 
\Phi_{{\bf r}+ \eta_{\bf r}{\bf e}_{\mu}}^{\dagger} 
\Phi_{{\bf r}+\eta_{\bf r} {\bf e}_{\nu}}^{} 
\nonumber
\\
&& - \frac{h s}{2} \sum_{ \langle {\bf r}{\bf r}' \rangle} 
(\hat{n} \cdot \hat{z}_i ) 
( \Phi^{\dagger}_{\bf r} \Phi^{}_{{\bf r}'} + h.c.) .
\end{eqnarray}
It is convenient to introduce a rotor variable $\phi_{\bf r}$ such that 
$[\phi_{\bf r}, Q_{{\bf r}'}] = i \delta_{{\bf r}{\bf r}'}$. Then we 
have $\Phi_{\bf r} = e^{- i \phi_{\bf r}}$ and $|\Phi_{\bf r}| = 1$. 
After such a transformation, the electric charge density $Q_{\bf r}$ 
can take any integer value. This approximation is legitimate
since the weight with large $Q_{\bf r}$ is suppressed by the 
antiferromagnetic $J_y$. 
We further carry out the standard procedure and implement a coherence
state path integral for the phase rotor variable. We integrate out 
$Q_{\bf r}$ and obtain the partition function
\begin{eqnarray}
\mathcal{Z} &=& 
\int \mathcal{D} \Phi^{\dagger} \mathcal{D} \Phi \mathcal{D} \lambda \,
e^{-{\mathcal S} - \sum_{\bf r} \int d\tau \lambda_{\bf r} ( |\Phi_{\bf r}|^2 -1)},
\end{eqnarray}
where the effective action $\mathcal{S}$ is given by 
\begin{eqnarray}
{\mathcal S} &=& \int d\tau \sum_{\bf r} 
\frac{| \partial_{\tau} \Phi_{\bf r}|^2}{2J_y}  - J_{\pm} s^2 
                      \sum_{{\bf r}} \sum_{\mu \neq \nu} 
\Phi_{{\bf r}+ \eta_{\bf r}{\bf e}_{\mu}}^{\dagger} 
\Phi_{{\bf r}+\eta_{\bf r} {\bf e}_{\nu}}^{} 
\nonumber
\\
&& - \frac{h s}{2} \sum_{ \langle {\bf r}{\bf r}' \rangle} 
(\hat{n} \cdot \hat{z}_i ) 
( \Phi^{\dagger}_{\bf r} \Phi^{}_{{\bf r}'} + h.c.),
\end{eqnarray}
and $\lambda_{\bf r}$ is introduced to impose  
the unimodular constraint $|\Phi_{\bf r}| = 1$. 
With a uniform saddle point approximation by setting 
$\lambda_{\bf r} = \lambda$, we obtain two spinon dispersions, 
\begin{eqnarray}
\omega_1 ({\bf k}) &=& \big[ 2 J_y ( \lambda - J_{\pm} L_1({\bf k}) + h |L_2 ({\bf k})| )\big]^{1/2},
\\
\omega_2 ({\bf k}) &=& \big[ 2 J_y ( \lambda - J_{\pm} L_1({\bf k}) - h |L_2 ({\bf k})| )\big]^{1/2},
\end{eqnarray}
where 
\begin{eqnarray}
L_1({\bf k}) &=& s^2 \sum_{i=1}^{12} \cos ({\bf k}\cdot {\bf a}_i) ,
\\
L_2 ({\bf k}) &=& \frac{s}{2} \sum_{\mu=0}^3 (\hat{z}_{\mu} \cdot \hat{n}) 
              \, e^{i {\bf k} \cdot {\bf e}_{\mu}} .
\end{eqnarray}
Here $\{ {\bf a}_i \}$ are twelve second-neighbor vectors of the diamond lattice.  
The parameter $\lambda$ is solved by the self-consistent equation 
$\langle \Phi^{\dagger}_{\bf r} \Phi^{}_{\bf r} \rangle  =1 $ with 
\begin{eqnarray}
\sum_{\bf k} [\frac{J_y}{\omega_1 ({\bf k})} + \frac{J_y}{\omega_2 ({\bf k})} ]= 2. 
\end{eqnarray}

\section{V. Distinction between the dipolar U(1) QSLs for DO doublets and usual Kramers' doublets}

Here we explain the difference between the dipolar U(1) QSL 
for DO doublets and the dipolar U(1) QSL for the usual Kramers' 
doublets. For the usual Kramers' doublets, the generic exchange Hamiltonian is~\cite{Savary2012,Sungbin2012,PhysRevB.78.094418,Onoda2011}
\begin{eqnarray}
H_{\text{Kramers}} &=& \sum_{\langle ij \rangle} J_{zz} S^z_i S^z_j 
                          - J_{\pm} (S^+_i S^-_j + S^-_i S^+_j) 
\nonumber \\
&+&  J_{\pm\pm} [ \gamma_{ij} S^+_i S^+_j + \gamma_{ij}^{\ast} S^-_i S^-_j ]
\nonumber \\
&+&  J_{z\pm} [ S_i^z ( \zeta_{ij} S_j^+ +  \zeta_{ij}^{\ast} S_j^- ) 
+ (i \leftrightarrow j)  ],
\label{eqkramers}
\end{eqnarray}
where $\gamma_{ij}$ is bond dependent phase factor 
that takes $1, e^{i2\pi/3}, e^{-i2\pi/3}$ on different bonds, 
$\zeta_{ij} = - \gamma^{\ast}_{ij}$, and $S^{\pm}_i = S^x_i \pm i S^y_i$. 
Please note the difference of $S^{\pm}$ from the definitiion of $\tau^{\pm}$ 
in the main text. In the parameter regime with $J_{zz} \gg |J_{\pm}|, |J_{\pm\pm}|, 
|J_{z\pm}|$ and the neighboring parameter regime,
the ground state of $H_{\text{Kramers}}$ is the dipolar    
U(1) QSL where the the Ising component $S^z$ behaves as the emergent 
electric field and the transverse components $S^{\pm}$ create
spinon excitations. All the spin components of an usual Kramers' 
doublet are magnetic dipole moments, thus all of them couple 
linearly with the external magnetic field and the neutron spin. 
Therefore, the inelastic neutron scattering detects 
both the gapped spinon continuum and the gapless gauge phonon
in the dipolar U(1) QSL for the usual Kramers doublets. 

For the DO doublet, the generic model is given by $H_{\text{DO}}$.
This model can be obtained from $H_{\text{Kramers}}$ if one simply sets
$\gamma_{ij}$ and $\zeta_{ij}$ to 1 on every bond, but the ground
states of $H_{\text{DO}}$ cannot be obtained from $H_{\text{Kramers}}$
in this manner. As we have described in the main text, 
what we have done is to perform a rotation about the 
$y$ axis in the pseudospin space to eliminate the 
crossing term $J_{xz}$. The resulting model is the XYZ model.

Let us here focus on the dipolar U(1) QSL in the regime $\tilde{J_z} \gg 
|\tilde{J}_x|, |\tilde{J}_y|$. In this phase, $\tilde{\tau}^z$ is the
emergent electric field and $\tilde{\tau}^x$ creates the spinon excitations. 
The external magnetic field and the neutron spin couple linearly   
to the $\tau^z$ component. Since $\tau^z$ is a combination of 
$\tilde{\tau}^x$ and $\tilde{\tau}^z$, the external magnetic field 
and the neutron spin couple with both the emergent electric field
and the spinons. For this reason, the inelastic neutron scattering
measurement detects both the gapless gauge photon and 
the gapped spinon continuum. This is clearly 
different from the origin for the usual Kramers' doublets.

\end{document}